# Multi-classification of High-Frequency Oscillations Using iEEG Signals and Deep Learning Models

Zayneb Sadek [1*], Abir Hadriche [1,2], Rahma Maalej [1,3], Nawel Jmail [1,3]

[1] Digital Research Center of Sfax, Tunisia
[2] Regim Lab, ENIS, Sfax University, Tunisia
[3] Miracl Lab, Sfax University, Tunisia
Email: zayneb.sadek123@gmail.com ; Abir.hadriche.tn@ieee.org; rahmamaalej1234@gmail.com
;naweljmail@yahoo.fr
*Corresponding author

*Abstract*—Over the past decade, high-frequency oscillations (HFOs) have been studied as a promising biomarker for localizing epileptogenic areas in drug-resistant patients requiring pre-surgical intervention, while exploiting intracranial electroencephalographic iEEG. Consequently, it's important to develop accurate methods for predicting epileptic seizures. Seizure prediction involves classifying appropriate indicators, which is a difficult classification problem. Deep learning techniques, such as convolutional neural networks (CNN), have shown great promise in analyzing and classifying epilepsy-related iEEG signals. In this study, we proposed three global methods, multiclass SVM, multiple architecture CNN, and CNN-SVM, which are evaluated on a simulated iEEG dataset and then on a real iEEG signal. Our best results for the three models yield high accuracy rates, GoogLeNet-SVM achieves approximately 99.63% and 94.07% for simulated data (1) and real data (2), respectively, SVM multiclass achieves 98.14% and 88.51% for (1) and (2), respectively, and GoogLeNet achieves 98.52% and 91.85% for (1) and (2), respectively. Furthermore, the proposed model performs better than other current techniques.
These results suggest that deep learning models could be a successful strategy for classifying epilepsy indicators and could potentially improve seizure prediction methods, thus improving the life quality of epileptic patients.

## I. Introduction

Epilepsy is the fourth prevalent neurological condition, which is characterized by unpredictable seizures. It affects individuals of all ages, including numerous patients, and is considered medically incurable and necessitates corresponding neurosurgery in order to achieve seizure freedom [1].
Preoperatively identifying epileptogenic zones is crucial for epilepsy surgery [1]. Presurgical biomarkers are therefore crucial in determining the regions known as epileptogenic zones, which cause epileptic seizures. Although the epileptogenic zone is difficult to measure and localize using inadequately concordant or inconclusive data from numerous tests, intracranial electroencephalography (iEEG), technologies [2, 3] are widely adopted.
iEEG is recorded as a traumatic technique where cortical sub-zone interactions are visualized by placing electrodes directly on the brain [4, 5].
Safe ablation of the EZ area and accurate localization are major factors for a good surgical outcome [6]. IEEG recordings are used to detect epileptogenic regions thanks to their ability to directly record epileptogenic discharges with high temporal and spatial accuracy and are regarded as the gold standard in electrophysiology for defining SOZ, which defines EZ to a large extent [7]. Electroencephalography (iEEG) recording using depth electrodes has become more popular in recent years.
since, it is less invasive than the subdural grid electrode technique [8, 9]. Moreover, epileptologists mainly focus on the critical iEEG to reveal the SOZ; However, intercortical HFOs have been considered a promising biomarker for localization of the epileptogenic zone [10, 11]. It has been well demonstrated and reproduced that HFO levels are higher in SOZ than outside [11-13]. However, HFOs are characterized as a transient phenomenon with weak and rapid oscillations, typically elapsing from 6 to 30 ms with a varied morphometry [14] Consequently, it is well recognized that manually identifying HFOs can be extremely laborious, time-consuming, and subject to subjective biases [15, 16].
 In this context, several algorithms for automated detection have been built, and implemented to help significantly minimize the amount of labor needed for HFO analysis and avoid biases induced by human evaluators [20]. However, the majority of HFO detection algorithms have been accomplished by simply thresholding instantaneous frequency traces, which could be vulnerable artifact influence and irregular HFO morphometry [18].
In the clinical study of HFOs, detection accuracy is quite important, yet, classifying the different events is also crucial. HFOs can be classified according to their frequency range into ripples (80-250 Hz) and fast ripples

(250-500 Hz) [19]. Fast ripples (FR) are thought to be more focal and closely related to epileptogenicity than Ripples (R) [20,21]. Evidence indicates that peak concomitant HFOs are more closely related to SOZ [22]. Consequently, a detection and classification framework were proposed that achieved high accuracy, sensitivity, and specificity.

Many studies in the literature aim to precisely categorize these diseases, using machine learning and deep learning models in order to classify HFO and promote seizure detection and epilepsy diagnosis.

The work of Daniel Lachner-Piza et al. [23] proposed a supervised machine learning approach that used SVM multi-classification of HFOs into 4 classes named Rs, FRs, IESs co-occurring during Rs, and IESs co-occurring during FRs.

Sciaraffa et al. [24] proposed a supervised machine learning approach using SVM and LDA. Logic regression KNN Multi-classification of HFOs into three classes: Rs, FRs, and FRs co-occurring during Rs.

Justin A. Blanco et al. [25] proposed unsupervised k-medoids for separation between HFOs (Rs, FRs, and Rs+FRs) and artifacts. Firpi et al. [26] proposed a supervised neural network to distinguish HFOs from baseline activity.

The work of M. Dümpelmann et al. [27] presented an HFO detector using a radial basis function neural network. The input features of the detector were energy, line length, and instantaneous frequency. Visually marked "ripple" HFOs (80–250 Hz) of 3 patients were used to train the neural network, and a further 8 patients served for the detector evaluation. Also, S. Chaibi et al. [28] proposed a Decision Tree to classify two classes (HFO and No-HFO). It has a sensitivity of 66.96%, and detects six features correlated with energy and duration.

Jrad N et al. [29] used the multiclass LDA method to classify (Ripples, Fast ripples, Ripple+Fast ripples, and artifact). It has a median of 80.5%, which energy was computed with a discrete wavelet.

As for Fatma krikid et al. [30], they proposed two approaches for the multi-classification of HFOs based on TF analysis. The first approach was a DL based on combining certain characteristics extracted from the TF representation of HFO with TF associated images binarized. It is divided into four frequency bands: gamma ([30, 80] Hz), high gamma ([80,120] Hz), Ripples (RS [120, 150] Hz), and Fast Ripples (FRs [250, 500] Hz). This coupling aims to provide a complete characterization of HFO. A second approach focuses on providing an automatic multi-classification method for HFOs based on CNN. They increased the data base using augmented reality in order to generate new TF images and evaluate its impact on CNN model performance.

The proposed algorithm ought to have robust classification capability to identify HFO subtypes and HFOs concomitant with other intercortical epileptiform discharges.

In our work, sections are arranged as follows: in Section 2, we present a description of our proposed global system with data collection and experimental methodology. In Section 3, we briefly provide experimental results, and in Section 4, we further discuss the obtained results. In Section 5, we conclude and offer suggestions for more work.

## II. METHODS AND MATERIALS

This section presents the pipeline of our proposed approach. This architecture presents the steps for detecting HFOs and automatic labeling three clusters for the dataset (1) (R [80 250] Hz, FR [250 500]) Hz, and SR (spike-ripple) and three clusters for the dataset (2) (R, FR, and R and FR). The steps of our proposed method are illustrated in Fig.1:

First, iEEG data are mapped into 2D time-frequency plots using continuous wavelet transform (CWT). Then, we improved model development with multiclass SVM architectures, model CNN, and CNN-SVM.

Performance is measured by the following evaluation metrics: the confusion matrix, accuracy, specificity, sensitivity, and F-measure of the proposed system. Following this, we validated our algorithm in a real-world setting, exploring the same previous strategies.

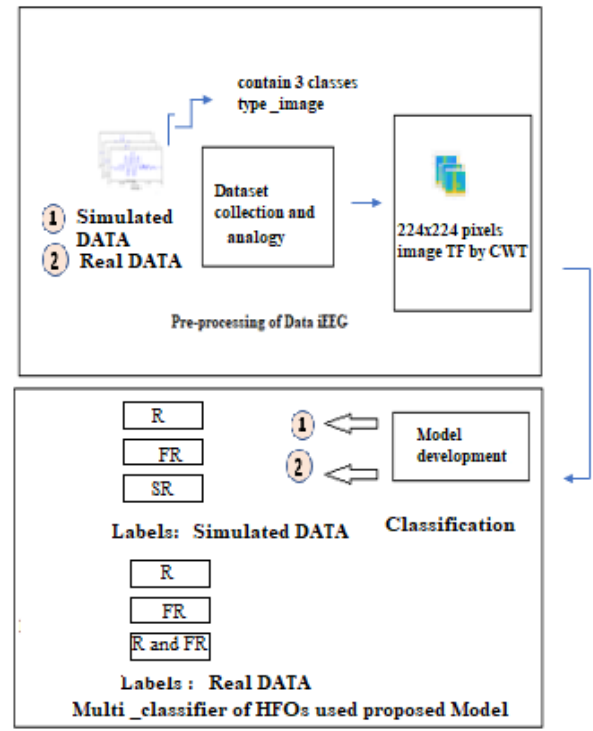

Fig.1. Proposed model for the classification of HFOs

### A. Dataset description

In this paper, two types of databases have been explored to assess the effectiveness of the suggested approaches:

**Simulated data** is obtained by a combination of a peak and HFO shapes as a real iEEG signal, sampled at 512 Hz and of duration 2s, with 1024 samples [31].

We have three classes of signals prepared, consisting of HFO (Ripple, Fast Ripple) and superimposed ripple and spike, thanks to various tests. Through altering various parameters, such as relative amplitudes, frequency of oscillations, signal-to-noise ratio (SNR), and overlap rate. We obtained 3000 data sets composed of HFO events (Ripple and Fast Ripple), and a mixture of spikes and fast events. We also varied the oscillation frequency within this range [85, 105, 200, 350, 450] Hz (ripples and fast ripples). **Real data** is obtained in combination with two sources. First, iEEG recordings for a patient with pharmacoresistant epilepsy, where the clinical neurophysiology department of La Timone Hospital in Marseille was responsible for the acquisition and preparatory measures [32] and validated by an expert neurologist were recorded on a Deltamed system, sampled at 1000 Hz with a low-pass filter.

This particular iEEG signal has a frame rate of 300 ms and consists of 117 canal x 30 segments. We have chosen the first 1000 data of the ripple. The second dataset is iEEG data with marked HFO events [33] sampled at 2000 Hz. We included 20 patients, of whom 9 had mesial temporal lobe epilepsy (TLE) and 11 had extratemporal epilepsy (ETE). For each patient 28 intervals, we recorded intervals with 300 ms of frames, and we explored in total 2220 trials of HFO.

### B. Data Preprocessing

The most relevant feature of HFOs for visual evaluation is the localized power distribution in frequency and time domains, which is viewable in two dimensions (2D) time-frequency using a continuous wavelet transform (CWT). The scalogram image is represented by the *cwt* coefficients of the iEEG data, Various wavelets have been suggested for various data processing applications. In time-frequency analysis, wavelet analysis can be thought of as a windowed Fourier analysis with a variable window width. They can provide details on an event's local frequency structure.

*cwt* is a convolution of the wavelet function $\psi(t)$ with the signal $x(t)$ and is given by Eq. (1) [34].

$$cwt(b,k) = \frac{1}{\sqrt{|k|}} \int_{-\infty}^{+\infty} x(t) \psi\left(\frac{t-b}{k}\right) dt \quad (1)$$

where the translation parameter is denoted by $b$ and the scaling parameter by $k$. Since $k$ and $b$ are continuous parameters, many wavelet coefficients are produced. The wavelet used for CWT is the Analytic Morlet, which has equal variance in frequency and time. Wavelets that are complex-valued in the time domain and have one-sided spectra are known as analytical wavelets. These parameter values allowed us to convert a 1D iEEG signal dataset into a scalogram image dataset. The scalogram images were scaled to dimensions of 224× 224×3 pixels as an input image for the proposed model and ranged horizontally for 300 ms and vertically from 80 to 500 Hz.

### C. Model Development

In this work, we are interested in the creation of our model, and we start by applying machine learning techniques.

*1) Multiclass SVM*

To explore this method for multi-class classifications, two main approaches have been suggested, based on reducing multi-class to a set of binary problems.

The initial strategy is known as one-versus-all, in which a set of binary classifiers is trained to distinguish between each class and the others. Then, based on the highest decision value, each data object is assigned to a class [35]. This method results in N-SVM (where $N$ is the number of classes) with $N$ decision functions. Although this is a fast method, it has errors that are caused by slightly unbalanced training sets. The second approach is called "one-on-one." In this, a series of classifiers are applied to each pair of classes, with the most calculated class kept for each object. Next, the max-win operator is employed to determine which class the object will ultimately be assigned to. This method requires the application of $N(N-1)/2$ machines. Compared to the "one-against-all" approach, this one requires more computation. It was found to be more suitable for multi-class cases.

**Error-correcting-output-codes (ECOC) for multiclass SVM classification**

The concept of the multiclass method is based on error correction output coding (ECOC). It consists of applying binary (two-class) classifiers to solve multiclass classification problems.

This approach is based on converting the $M$ class classification problem into a large number $L$.

ECOC assigns a unique codeword to a class instead of assigning a label to each class. An error correction code $(L, M, d)$ is $L$ bits long, with $C$ a single codeword with a hamming distance $d$. The hamming distance between two codewords presents the difference in the number of bit positions. In a classification problem, $M$ is the number of classes and $L$ is a number determined by the method used to generate error correcting codes.

A variety of techniques are suggested, including BCH codes [36], and exhaustive codes [37] are proposed to produce error corrections codes. [38] proposed using codes with a maximum Hamming distance between them and suggested that errors $(d-1)/2$ can be corrected in codewords for a Hamming distance d between codes.

Decomposition of a $c$ class multiclass problem having $k_1, \ldots, k_c$ as the class labels generated a set of $m$ binary classifiers represented by $f_1, \ldots, f_m$ a binary classifier subdivides the input patterns into two complementary super class $k_i^1$ and $k_i^{-1}$ grouping together one or more classes of multiclass problem.

Let $M = [b_{ij}]$ is a decomposition matrix of dimension mxc, connecting class $k_i, \ldots, k_c$ to the super classes $k_i^1$ and $k_i^{-1}$, where an element of matrix $M$ can be defined as:

$$b_{ij} = \begin{cases} 1 & \text{if } k_c \subset k_i^1 \\ -1 & \text{if } k_c \subset k_i^{-1} \end{cases} \quad (2)$$

Therefore, for $M$ classes, according to matrix $D \subset \{\pm 1\}^{M \times C}$ is obtained.

*2) Convolutional Neural Network (CNN)*

Convolutional neural network is a specific type of multi-layer neural network, a simple neural network cannot learn complex features. In multiple applications [39, 40] such as image classification, object detection, and medical image analysis, CNNs have shown excellent performance.

In this part of the work, several convolutional neural networks have been suggested and proposed, such as GoogLeNet, ResNet 18, ResNet 50, and ResNet 101, which were tested on different CNNs. It is a multi-layer network structure that basically consists of five layers, starting from the input layer, convolutional layer, pooling layer, fully connected layer, and finally the output layer, as shown in Fig. 2.

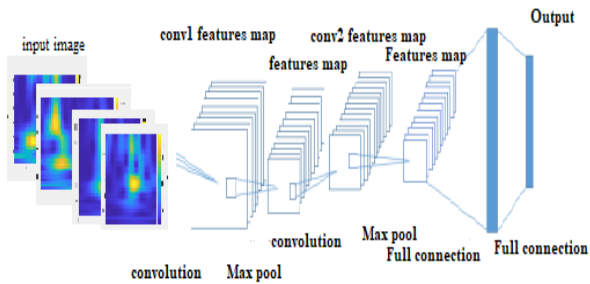

Fig.2 The basic structure of CNN

With CNN, we can transfer local features from inputs at higher layers to lower layers for greater complexity and functionalities. Therefore, we propose a more detailed description of CNN:

- Convolution layer: is the main layer of CNN. It performs a process called "convolution", which is a linear process that involves multiplying input-set weights. Two sets of information are combined by performing a mathematical operation between the kernel and the input image.

- Pooling layer: usually comes in after the convolutional layer. Pooling is a simple operation where squares of pixels (usually $2\times 2$ or $3\times 3$) are substituted with a single value, effectively reducing and simplifying image size. In the CNN model, pooling can be performed using maximum, average, or sum pooling. Because maximum pooling makes it possible to identify more exact features, it was used in this study.

- Fully Connected ($f_c$) Layer Also known as the dense layer, the CNN architecture ends with the fully connected layer, which resembles a standard neural system.

- Last layer activation function: softmax

The activation function used for the final fully connected layer differs from that of the other layers. Each task requires the selection of an appropriate activation function. The softmax function is an activation function used in the multiclass classification task that normalizes output real values from the last fully connected layer to target class probabilities, where each value ranges between $0$ and $1$ and all values sum to $1$.

This study has employed four distinct convolutional neural network techniques, namely:

GoogLeNet

GoogLeNet, also called Inception, is a deep convolutional neural network architecture design created By Google researchers [41]. Basically, it is 22 layers deep. GoogLeNet stands out for its in-depth architecture and efficient use of computing resources. He introduced the concept of build modules, which are modules containing multiple parallel convolutional operations of different kernel sizes. These parallel operations capture information at different scales, allowing the network to efficiently learn hierarchical features. The network has an input image size of 224 by 224. Our experimental study has made use of it. for 144 layers, and 6.9 million of its parameters are trainable. This network starts with the input layer, convolution layer, ReLU, and Max pooling. This structure is repeated until it reaches the fully connected layer, then the classifier layer occurs to give the output classification result.

The architecture addressed the challenge of vanishing gradients in very deep networks and achieved impressive performance in image classification tasks, particularly in the ImageNet Large Scale Visual Recognition Challenge (ILSVRC) in 2014. GoogLeNet demonstrated the ability to achieve competitive accuracy with significantly fewer parameters compared to others. in other deep networks of his time.

ResNet

ResNet (Residual Network) is a deep convolutional neural network architecture designed to address the challenges of training very deep neural networks. The key innovation of ResNet is the introduction of residual learning blocks, also called residual units or residual blocks. These blocks have shortcut connections that allow the model to get around one or more layers to learn the residual mappings. The main advantage of residual connections is that they facilitate the formation of very deep networks by alleviating the vanishing gradient problem. With residual connections, even if the information from the initial layers is not very informative, it can still be directly propagated to later layers, making it easier for the network to learn complex mappings.

ResNet architectures have been widely adopted in various computer vision tasks, such as image classification, object detection, and segmentation, due to their ability to efficiently train and scale to very deep networks.

- **ResNet50** is a specific variant of the Residual Network (ResNet) architecture. This method is characterized by its depth, consisting of 50 layers in total and 25.5 million parameters.
  The key innovation of ResNet-50, as in other ResNet architectures, is the use of residual learning blocks. These blocks contain shortcut connections, also called skipped connections, that allow the model to learn the residual mappings. This helps alleviate the vanishing gradient problem and makes it easier to train very deep neural networks.

The ResNet-50 architecture includes several building blocks, including residual units, which are coherent into a series of convolutional layers with batch normalization and rectified linear unit (ReLU) activations. Skipped connections in these units allow the direct flow of information from one layer to another, leading to the performance degradation that can occur in extremely deep networks.

- **ResNet 18:** The number of layers is 18, because 18 tells us about the layer of the network. The total number of parameters is 11,511,784, and all the parameters are trainable [42]. The network design used in this research has 72 layers. ResNet18 is composed of $3 \times 3$ CONV layers with filters. There are only two pooling layers used at the beginning and end of the network, and identity connections are found between every two CONV layers. ResNet18 uses shortcut connections to solve the disappearance problem [41]. Additionally, the two pooling layers in the ResNet18 design are not countable.
- **ResNet101:** The Resnet-101 structure is made up of 101 couches. It is based on the residual neural network learning method. About 44.6 million of its parameters are trainable. The network depth is 347 layers, Resnet-101 differs from other architectures by optimizing the residuals between the desired input and convolution properties. The desired functionality is achieved more easily and efficiently compared to other architectures. Thus, a deeper network's parameter count can be decreased by using residual optimization by specifying the number of parameters, the number of canapes can be effectively reduced [43,44].

In ResNet architecture, information that cannot be learned in the previous layer is applied from the old layer to the new one with the ResBlock layer. The residual values are passed to the following layer in the Resnet architecture by the Resblock layer. At each two-layer activation, this jumps between the weight layers, and the Relu activation code adds value modifying, the system account.
The nonlinear Relu function is represented by the two layered residual block structure.

$$F = w_{2\sigma}(w_1 x) \qquad (3)$$

This result's y output is obtained by adding a second Relu value.

$$y = f(x, (\{w_i\})) + x \qquad (4)$$

In Eq.10 x denotes the input vector, while y denotes the output vector.

   *3) The Proposed CNN-SVM hybrid Architecture*

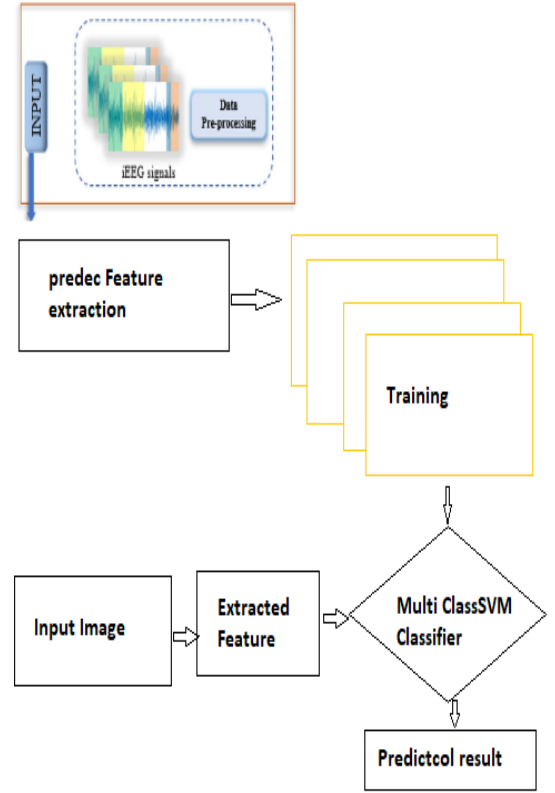

Fig.3. CNN_SVM model architecture

Multiple popular pre-trained networks, such as AlexNet, GoogLeNet, ResNet50, VGG-19, and others. These networks are trained on data set simulated and then real.
The input image size (224 x 224 x 3 pixels) is defined by the first layer.
A pre-trained CNN is used for feature extraction, with a starting network layer capturing the main image feature (edges, blobs). A convolutional layer contains network weight. A 2 D TF feature for pre-trained CNN extracts. The preparation of the training and test datasets is carried out by dividing the dataset into 70% training data and 30% testing data. Processing of the training and test datasets is carried out by the CNN model. An SVM classifier is trained by the features that are extracted using the CNN present in the training data set.
During the testing process, using the same model, test features are extracted and passed to the classifier to perform the tests.
**Step classification of HFO using the CNN_SVM model hybrid.**

**Step 1:** Convert signal 1D iEEG to 2D TF scalogram image
**Step 2:** Enter an image from one of the category folders.
**Step 3:** Load database: load images using an Image DataStore function that operates on the image location to hold the images and labels associated with each image category. An Image DataStore allows us to store voluminous image data, it also divides the data into 70% training data and 30% test data. The smallest number of images in each category is determined by each label. Then we tested the pre-trained ResNet 50, ResNet 101, ResNet 18, google NET, and VGG19 using the functions resnet50, resnet101, resnet18, googlenet, and vgg19, respectively.

**Step 4:** Image pre-processing: the CNN model processes both the training Set and the test Set. Image pre-processing for CNN according to the network used is carried out by resizing the image according to the network (224 by 224).

**Step 5:** Extract training features with CNN using the Activation function with the feature layer of each algorithm from resnet50, resnet18, resnet101, vgg19, and GoogLeNet.

**Step 6:** The training of a multiclass SVM classifier is completed using CNN's features.

**Step 7:** Evaluation of the classifier is performed by taking image features from the test dataset. Subsequently, the classifier receives these features back to determine how accurate the trained classifier is.

**Step 8:** Estimated classification: The feature utilized in the preceding actions can be found in
https://www.mathworks.com/examples/computer-vision/mw/vision-ex77068225-image-category-classification-using-deep learning

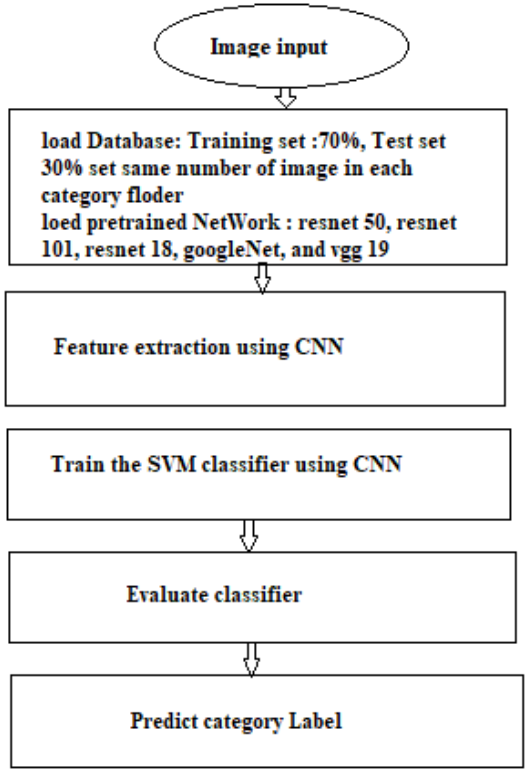

Fig.4. Model classification steps CNN_SVM

### D. Metrics

Measuring Classification Performance: accuracy, precision, F-measure, sensitivity, and specificity.

Performances of classification are computed in order to define the robustness of the proposed system,

In this study, we operated on a confusion matrix to compare classification results, using:
Accuracy formulated in Eq. (5).

$$Acc = \frac{TP + TN}{TP + FP + FN + TN} \quad (5)$$

Precision formulated in Eq. (6).

$$Precision = \frac{TP}{TP + FP} \quad (6)$$

F-Measure (F1-Score) presented in Eq. (7).

$$F1 - score = \frac{2 \times TP}{(2 \times TP + FP + FN)} \quad (7)$$

Sensitivity (SE) formulated in Eq. (8).

$$SE = \frac{TP}{TP + FN} \quad (8)$$

Specificity (SP) formulated in Eq. (9).

$$SP = 1 - \frac{FP}{FP + TN} \quad (9)$$

### III. EXPERIMENTAL RESULTS

The experiments in this study were performed on a PC equipped with a six-core Intel i7 processor using MATLAB (2023b). The server was equipped with an NVIDIA GEFORCE 920M with 6 GB of memory. This section presents the results of classifying HFO epilepsy indicators for iEEG signals using our models, SVM multiclass, multiple architecture CNN, and CNN-SVM. We conducted both methods on different data sets.

#### A. The Results Obtained by Applying SVM multiclass

In these experiments, we tested the results of SVM ECOC on different data sets. First, we employed two real and simulated data sets. To demonstrate multiclass image classification, the experiment was carried out with three categories taken from the simulated dataset (1) and three categories taken from the real dataset (2). Each category for data set (1) and data set (2) is assigned an index ranging from 1 to 3. A multiclass SVM is a classifier that predicts the label of categories with its index. The five performance measures evaluated are precision, F-measure, specificity, sensitivity, and accuracy.

The confusion matrix in Fig. 5 shows the performance measures evaluated for each category, respectively, for data set (1) and data set (2). The experimental results of the classification of images with predicted image labels and indexes are present. These experiments were evaluated after 10-fold cross-validation. Table I shows a summary of the results based on the test performed by SVM. Consider the datasets (1) and (2) with 224x224 image size.

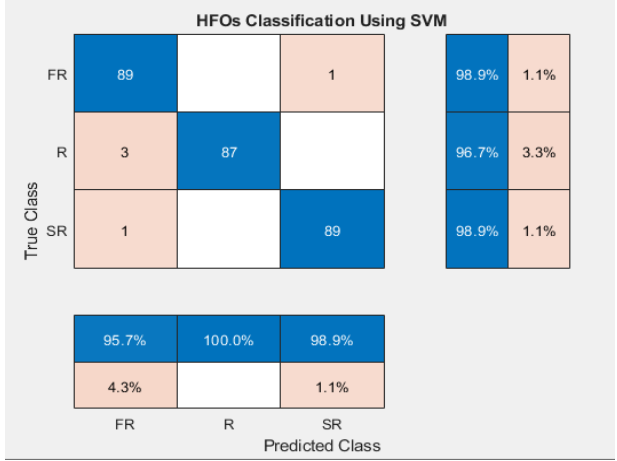

a)

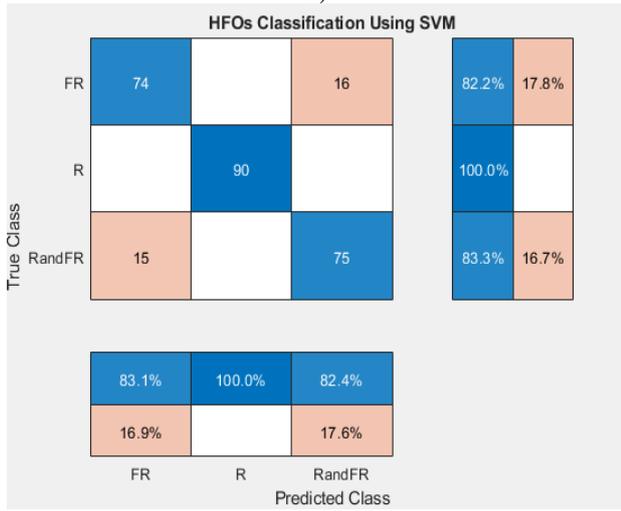

b)

Fig.5. Confusion matrix analyses based on SVM multiclass after 10-fold cross validation model obtained from a) data set (1), and b) data set (2).

TABLE I: PRECISION, $F$ MEASURE, SPECIFICITY, SENSITIVITY, AND ACCURACY MACRO AVG BY SVM ECOC FROM TWO DATA SETS.

conditions of learning rate 0.001 and Adam optimization function.

As can be seen in Fig. 6, the confusion matrix of the four classifiers. However, it provides information on the total number of correctly classified segments and the total number of misclassified segments. The first three rows in the confusion matrix table relate to the predicted category class (output class), and the true class (target class) is related to the first columns. The diagonal cells show how many categories have been appropriately classified. Cells outside the diagonal refer to categories that have been misclassified. The result shows that they are best suited for image classification.

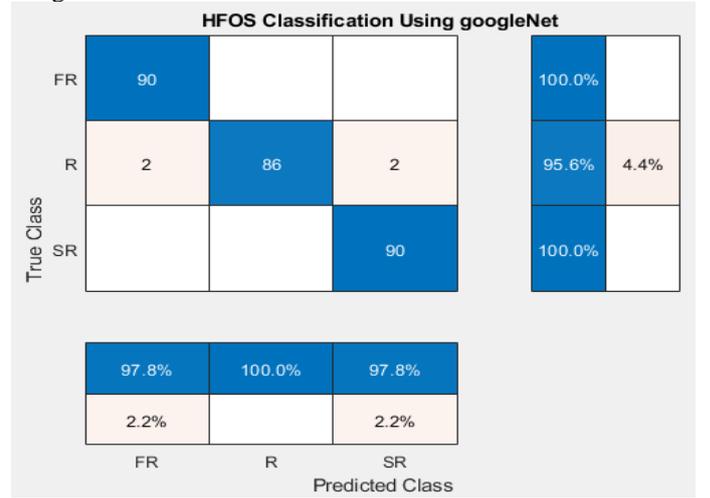

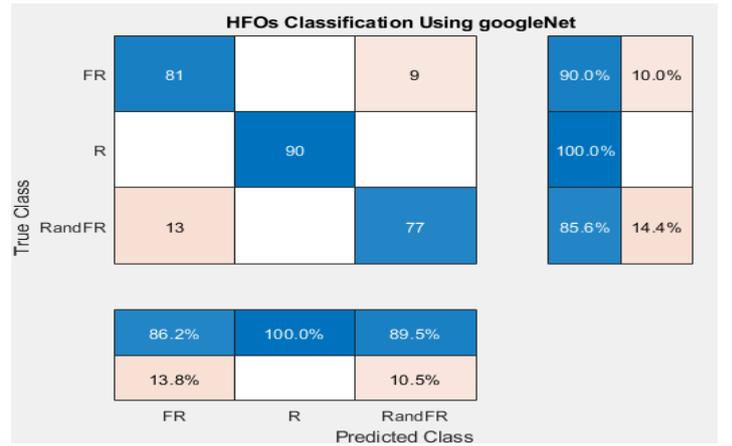

a)

| Model | Data set | Precision | F-Measure | Specificity | Sensitivity | Accuracy |
|---|---|---|---|---|---|---|
| **SVM Multiclass** | Simulated DATA (1) | 98.19 | 98.15 | 99.07 | 98.14 | 98.14 |
|  | Real DATA (2) | 88.52 | 88.51 | 94.25 | 88.51 | 88.51 |

## B. Results Obtained by Applying GoogLeNet, ResNet50, ResNet101, and ResNet18

Table II presents a summary of the results based on different tests performed. Consider the dataset with the same conditions for images with a size of 224×224. The GoogLeNet achieved the highest accuracy at the same

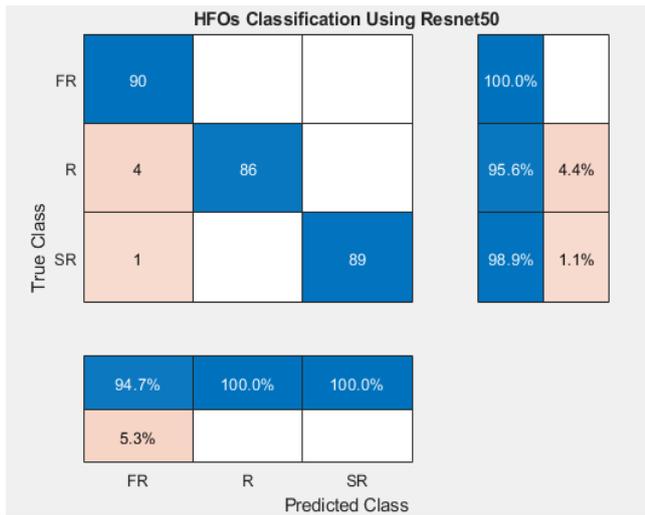

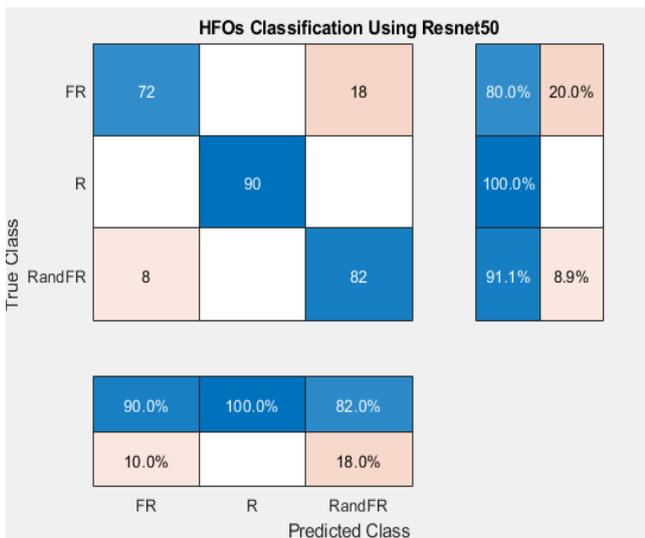

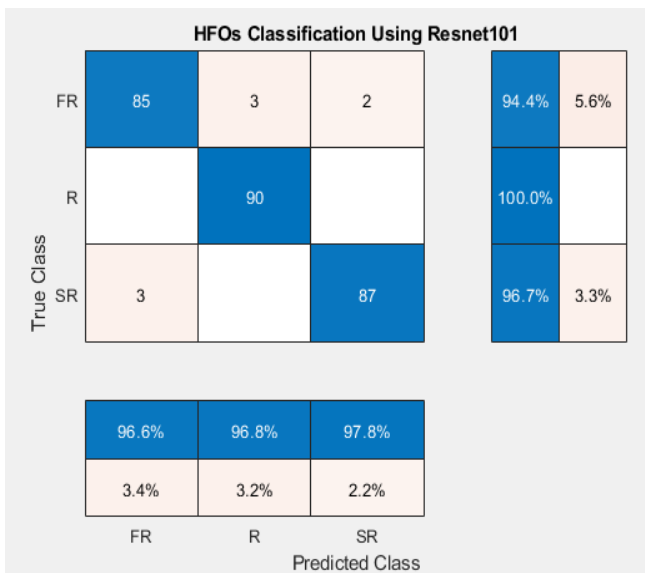

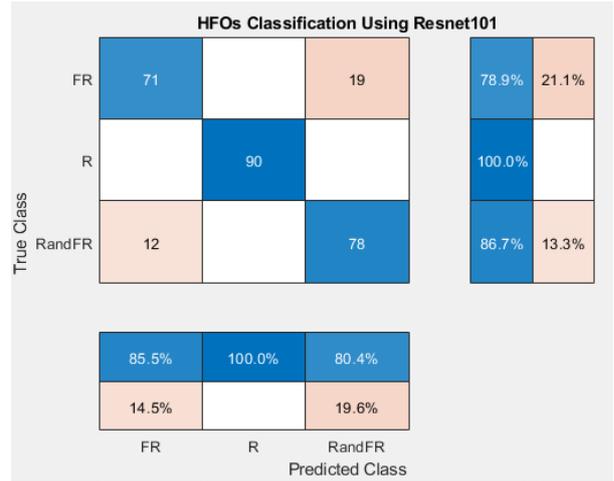

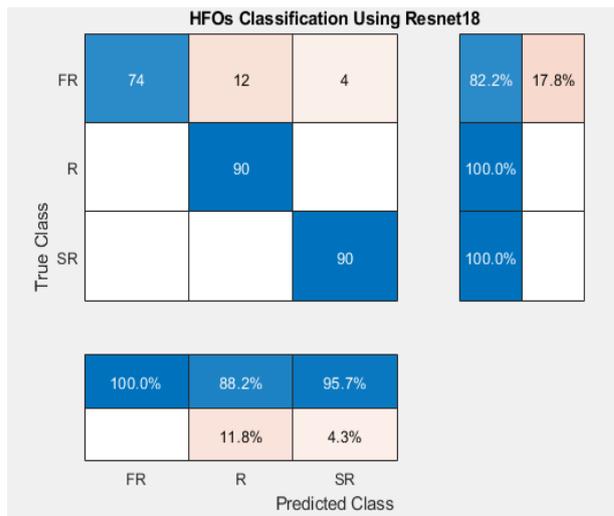

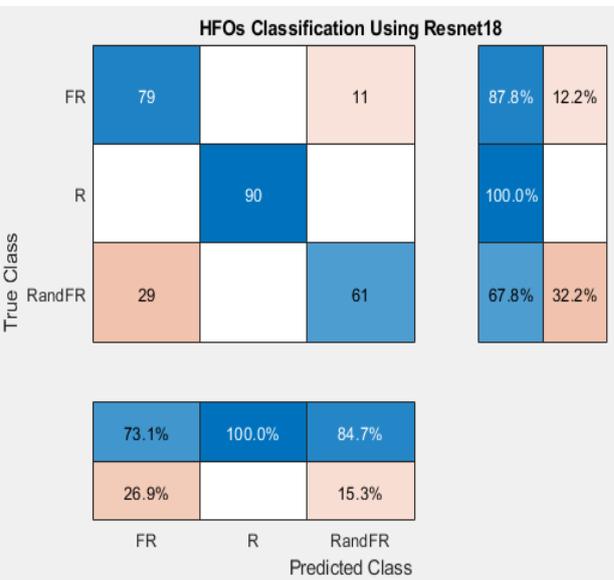

Fig.6. Confusion matrix of Test analyses based on: a) GoogleNet, b) ResNet50, c) ResNet101, and d) ResNet18 models from data sets (1), and data sets (2).

TABLE II. CLASSIFICATION PERFORMANCE COMPARISON BETWEEN FOUR MODELS CNN USING TWO DATA SETS

| Model | Data set | Precision % | F-Measure% | Specificity % | Sensitivity % | Accuracy % |
|---|---|---|---|---|---|---|
| **GoogleNet** | **Simulated DATA (1)** | **98.55** | **98.51** | **99.25** | **98.51** | **98.52** |
|  | **Real DATA (2)** | **91.90** | **91.84** | **95.92** | **91.85** | **91.85** |
| **ResNet 50** | Simulated DATA (1) | 98.24 | 98.15 | 99.07 | 98.14 | 98.15 |
|  | Real DATA (2) | 90.66 | 90.34 | 95.18 | 90.37 | 90.37 |
| **ResNet 101** | Simulated DATA (1) | 97.03 | 97.02 | 98.51 | 97.03 | 97.04 |
|  | Real DATA (2) | 88.65 | 88.50 | 94.25 | 88.51 | 88.52 |
| **ResNet 18** | Simulated DATA (1) | 94.66 | 93.94 | 97.03 | 94.07 | 96.04 |
|  | Real DATA (2) | 85.95 | 85.03 | 92.59 | 85.19 | 87.04 |

Bold values represent highest performance.

### A. The Results Obtained by Applying CNN-SVM hybrid models

The model is a combination of two supervised classification techniques, CNN and support vector machine (SVM).
The models used batch size 128 and optimized the loss function using the Adam optimizer.
To take temporal information into account, five hybrid models based on the above CNN and SVM classifier models were built, and their classification performance was tested on simulated data then on real data.

The confusion matrices of the CNN_SVM model are depicted in the following Fig. 7, where the three classes for the data set (1) represent Fast Ripple (FR), Ripple (R), and Spike Ripple (SR) respectively. And also, data set (2) represents Fast ripple (FR), Ripple (R), and Ripple and Fast Ripple (Rand FR), respectively.

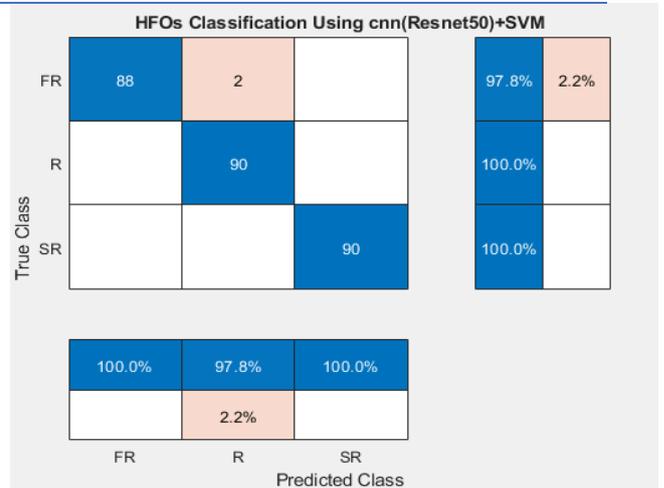

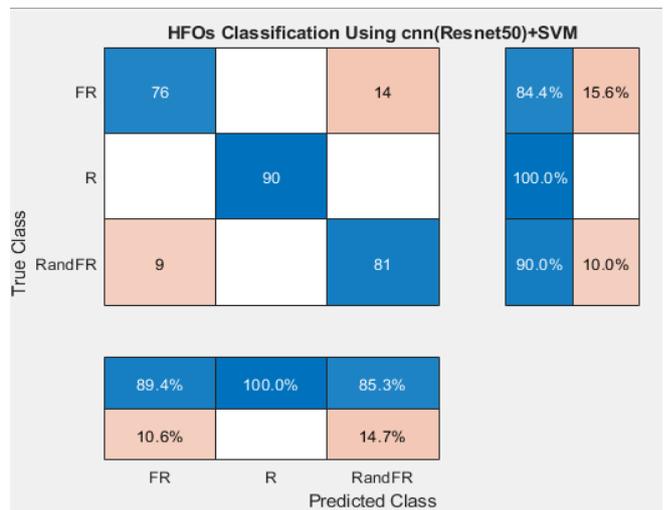

a)

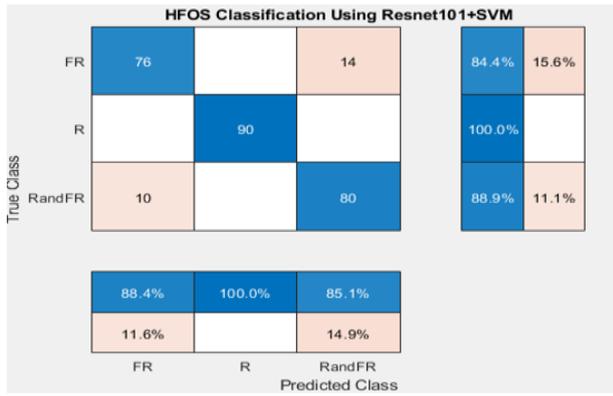

b)

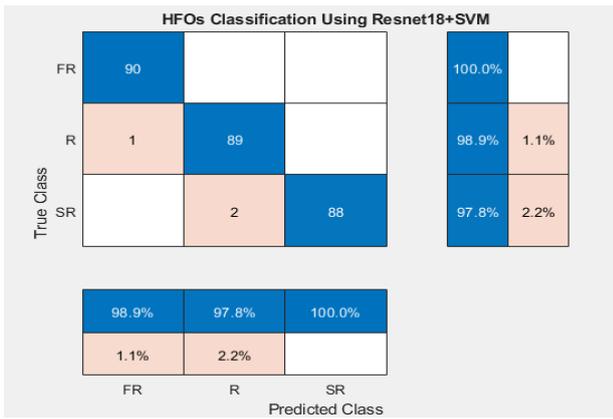

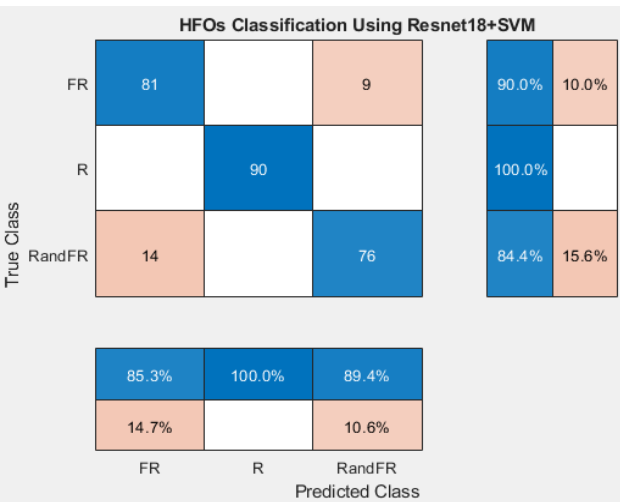

c)

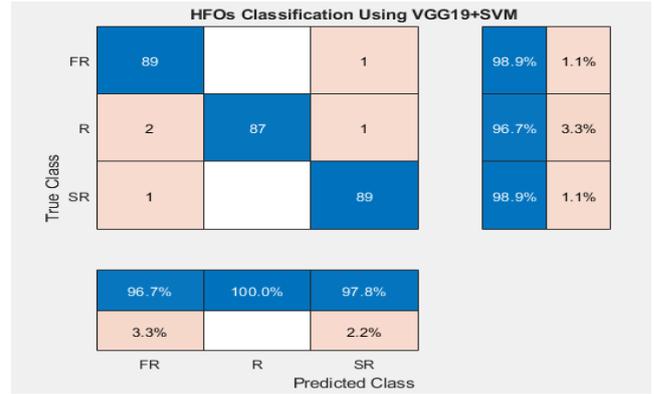

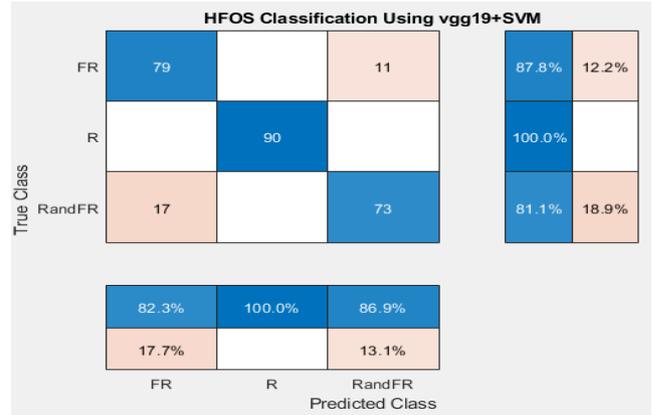

d)

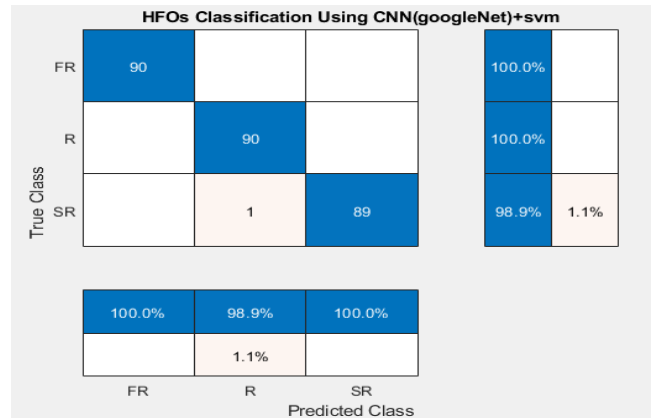

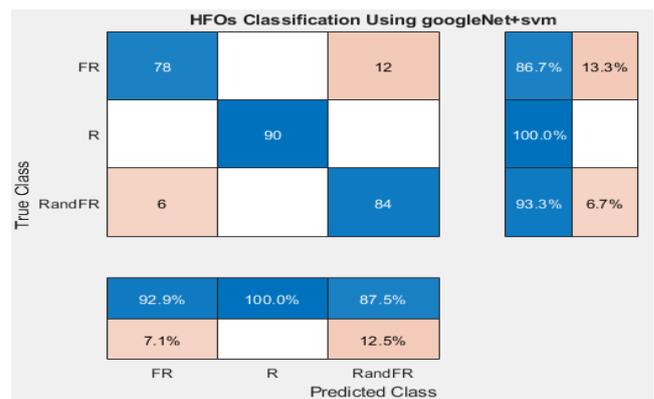

e)

| Model | Data set | Precision % | F-Measure% | Specificity % | Sensitivity% | Accuracy % |
|---|---|---|---|---|---|---|
| **Resnet50_ SVM** | **Simulated DATA [1]** | 99.27 | 99.25 | 99.63 | 99.25 | 99.26 |
| | **Real DATA [2]** | 91.55 | 91.47 | 95.74 | 91.48 | 91.48 |
| **ResNet 18_SVM** | **Simulated DATA [1]** | 98.90 | 98.99 | 99.44 | 98.88 | 98.89 |
| | **Real DATA [2]** | 91.55 | 91.47 | 95.74 | 91.48 | 91.48 |
| **ResNet 101_SVM** | **Simulated DATA [1]** | 97.87 | 97.77 | 98.88 | 97.77 | 98.52 |
| | **Real DATA [2]** | 91.12 | 91.11 | 95.55 | 91.11 | 91.11 |
| **GoogleNet _SVM** | **Simulated DATA [1]** | **99.64** | **99.63** | **99.81** | **99.63** | **99.63** |
| | **Real DATA [2]** | **94.08** | **94.07** | **96.66** | **94.07** | **94.07** |
| **VGG19_SVM** | **Simulated DATA [1]** | 98.18 | 98.15 | 99.07 | 98.14 | 98.15 |
| | **Real DATA [2]** | 89.73 | 89.61 | 94.81 | 89.63 | 89.63 |

Fig.7. confusion matrix of the test classification effects of the five CNN-SVM models: a) resnet50+svm for data sets(1), and data sets (2) respectively, b) resnet 101+SVM for data set (1), and data sets (2) respectively, c) resnet 18+SVM for data sets (1), and data sets (2) respectively, d)vgg19+svm for data sets (1), and data sets (2) respectively, and e) GoogLeNet for data sets(1), and data sets (2) respectively.

TABLE III. RESULT OF CLASSIFICATION OF DIFFERENT MODEL'S HYBRID CNN_SVM USING TWO DATA SETS.

Bold values represent highest performance

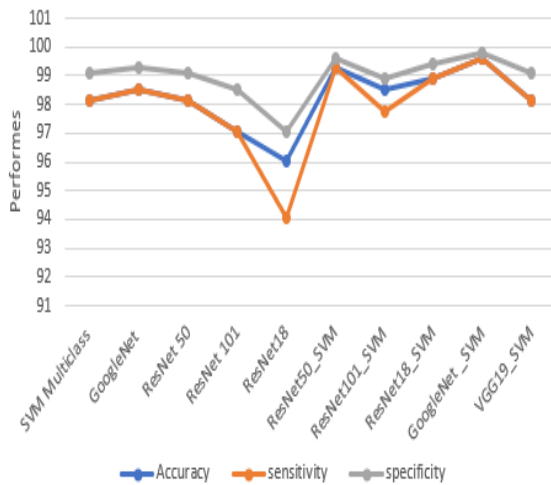

a)

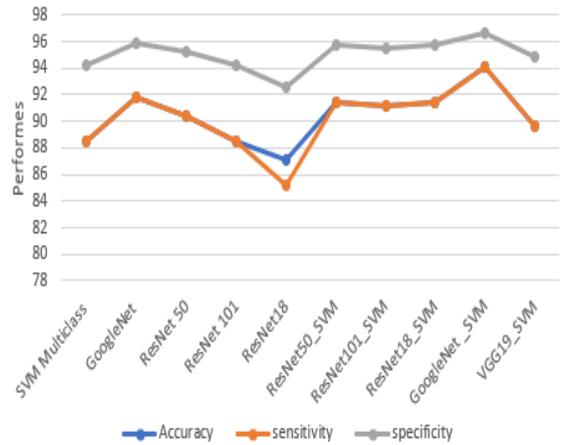

b)

Fig.8. Comparison graph of accuracy, sensitivity, and specificity between various test approaches used in this study on a) simulated data on the left, and b) real data on the right.

TABLE IV. PRECISION, F-MEASURE, SPECIFICITY, SENSITIVITY AND ACCURACY OF THE BEST RESULTS ACHIEVED FOR THREE MODELS

| Model | Data set | Precision % | F-Measure % | Specificity % | Sensitivity % | Accuracy % |
|---|---|---|---|---|---|---|
| SVM | Simulated DATA (1) | 98.19 | 98.15 | 99.07 | 98.14 | 98.14 |
|  | Real DATA (2) | 88.52 | 88.51 | 94.25 | 88.51 | 88.51 |
| GoogLeNet | Simulated DATA (1) | 98.55 | 98.51 | 99.25 | 98.51 | 98.52 |
|  | Real DATA (2) | 91.90 | 91.84 | 95.92 | 91.85 | 91.85 |
| GoogLeNet_SVM | Simulated DATA (1) | **99.64** | **99.63** | **99.81** | **99.63** | **99.63** |
|  | Real DATA (2) | **94.08** | **94.07** | **96.66** | **94.07** | **94.07** |

Bold values represent highest performance.

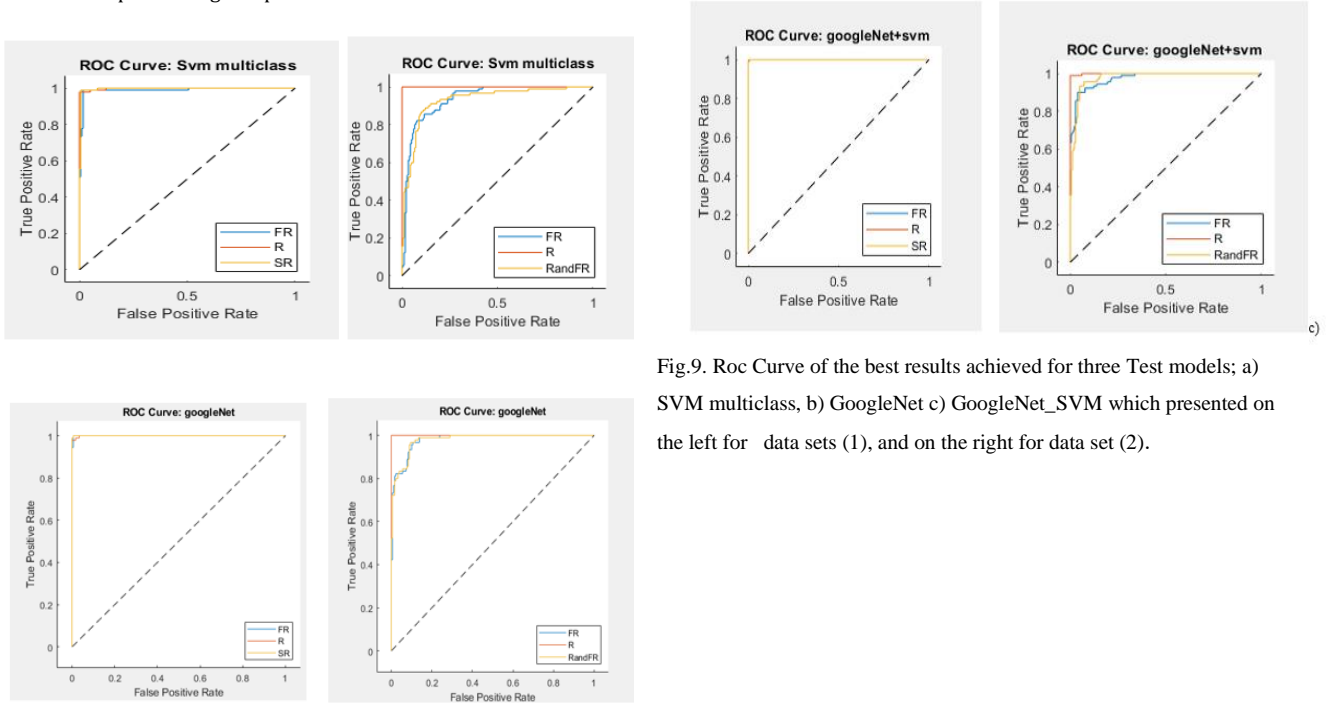

Fig.9. Roc Curve of the best results achieved for three Test models; a) SVM multiclass, b) GoogleNet c) GoogleNet_SVM which presented on the left for data sets (1), and on the right for data set (2).

IV. DISCUSSION

As mentioned, the main objective of this study was to classify HFO images into categories. The data set included 3000 images of simulated data and 3220 images of real data. The model CNN and hybrid CNN_SVM separated the data into training and testing sets. Then, they classified data into three classes: Fast Ripple (FR), Ripple (R), and Spike Ripple (SR) from data sets (1), R, FR, and FR and R for data sets (2), using the mean accuracy of training and test networks for 100 training.

The confusion matrix in Fig. 5, Fig. 6 and Fig. 7 shows the performance measures evaluated for each category for

| AI Technique | Methods | Context | Accuracy | Sensitivity | F1-score |
| --- | --- | --- | --- | --- | --- |
| M. Dümpelmann et al,2012 [27] | Supervised/RBF neural network | Discrimination between true and false HFOs | NA | 49.1% | NA |
| S. Chaibi et al,2014.[28] | Supervised/decision trees | Discrimination between HFOs and background activity | 82.08% | 66.96 % | NA |
| Jrad N et al,2015.[29] | Multiclass LDA | Multiple classification of HFOs into 5 classes named; (Ripple,Fast Ripple , Ripple +Fast Ripple and artifact ) | NA | Median 80.5% | NA |
| N. Sciaraffa, et al,2020.[24] | Supervised/SVM; LDA; logistic regression; KNN | Multiple classification of HFOs into 3 classes named: Rs, FRs and FRs co-occurring during Rs | NA | 87.4% | NA |
| Fatma krikid et al.,2023[30] | Support Vector Machine (SVM) based on Radial Basis Function (RBF) | Classified SOIs: gamma/High gamma_/Rs/FRs/IESs Test for simulated data | 99.5% | 99.0% | 99.0% |
| | | Test for real data | 90.6% | 76.5% | 76.8% |
| Proposed model | GoogLeNet -SVM | R, FR, SR For simulated data | **99.64%** | **99.63 %** | **99.63%** |
| | | R, FR, and Rand FR used real data | **94.08%** | **94.07%** | **94.07%** |

SVM, CNN models, and CNN_SVM, respectively. The experimental results of the image classification with the predicted Image label and Index are presented.

The following four tables represent the different experiment results. However, HFO detector was compared to machine learning, including deep learning model, respectively. Table I shows validation accuracy with 10-fold cross validation strategy using the Error-correcting-output-codes method obtained by two data sets. Following the successful results obtained from the confusion matrix, the sensitivity, specificity, accuracy, and F-measure values for the IEEG data that the SVM model predicted correctly and incorrectly were calculated according to the values in the cross-classification given in table I.

To evaluate the model accuracy, sensitivity, specificity, precision, and F measure the application was tested on multiple CNN network architectures with a multiclass SVM model on two datasets. In Table II, we can see that the GoogLeNet model consistently outperforms the three well-known CNN models used as comparisons, such as ResNet 50, ResNet 18, and ResNet101 on two sets.

In table III shows that the outstanding classification performance of the present research work was mainly achieved by utilizing pre-trained CNN with GoogLeNet network architecture. Therefore, classification tasks can produce good performance when using CNN as a feature extractor and an SVM classifier with ECOC framework for multiclass classification problems, that results of GoogLeNet combined with SVM perform better than the

TABLE V.  RESULTS ANALYSIS AND COMPARISON OF PREVIOUS WORKS.

∗NA: not available.
Bold values represent the highest performance

Other methods as validated in the graphs Fig 8(a) and Fig 8(b). In fig. 9 the best classifier is presented in the Roc curve. However, ROC curves are drawn for different subclasses in different datasets. The value of the area under the ROC curve that belongs to a value between 0 and 1 (a larger area indicates a better performance of the classifier). As can be seen in Fig. 9(c) the highest AUC is obtained by the GoogleNet-SVM classifier on two sets. In this test, the hypothesis is that the hybrid models respond very strongly thanks to superior characteristics.

In order to compare the performance results of the proposed GoogLeNet within the current application domain, we collected and compared findings from other recent studies conducted by various researchers, as presented in Table V. We found five relevant articles [27-29,24,30] in the literature that used different databases than the one used in this study.

M. Dümpelmann et al. [27,45] reported a supervised learning RBF neural network-based method for discriminating between true and false HFOs, which obtained a sensitivity of 49.1%. In [28], S. Chaibi et al. used a decision tree algorithm for discriminating between HFOs and background activity, achieving an accuracy of 82.08% and a sensitivity of 66.96%. Jrad N et al. [29] used multiclass LDA for classifying HFOs into (Ripple, Fast Ripple, Ripple-Fast Ripple, and artifact) with approximately sensitivity of 80.5%. N. Sciaraffa et al. 24] reported a machine learning algorithm based on SVM, LDA, logistic regression, and KNN for multi-classification of HFOs into 3 classes (Rs, FRs, and FRs co-occurring during Rs), with a sensitivity rate of 87.4%. Fatma Krikid et al. [30] introduced a method based on decomposition in Radial Basis Function (RBF) with Support Vector Machine (SVM) for the classification of gamma, high gamma, Rs, FRs, and IESs, achieving an accuracy of 90.6%, sensitivity of 76.6%, and F1-score of 76.8% obtained by real data. also, for simulated data, achieving an accuracy of 99.5%, sensitivity of 99.0%, and F1-score of 99.0%.

In this work, the main novelty is the implementation of a deep GoogleNet_SVM model for the automated classification of IEEG signals. A model is proposed as it provides good convergence and the highest performance accuracy, sensitivity, specificity, and F-measure. All parameters of the GoogLeNet_SVM structure are carefully fine-tuned in order to obtain a model with an optimal convergence rate, based on the experimental results, it can be concluded that the deep GoogLeNet features of iEEG time-frequency maps, combined with an SVM classifier, could serve as an applicable measurement for the accurate classification of HFO subclasses. However, deep networks allow more complex, non-linear functions to be learned, then, such networks can be difficult to converge. On the other hand, shallow network is easier to train but the features extracted are simple and may not be adequate for classification. Even though this proposed model could have the best classification performance as compared to the published works recorded in Table IV, the proposed GoogLeNet-SVM model still managed to obtain 99.63% accuracy, 99.63% sensitivity, and 90.81% specificity obtained by data sets (1), 94.07% accuracy, 94.07% sensitivity, and 96.66% specificity in data sets (2). This shows that given more iEEG data, the proposed model can achieve better results with pre-processing on 2 D TF map of the iEEG data. Thus, the overall diagnostic performance using the proposed model can be improved with more iEEG data.

## V. CONCLUSION

This study focuses on the prediction of indicators of pharmaco-resistant epileptic seizures using iEEG recordings via different applications of multiclass SVM, model CNN, and CNN-SVM. The CNN is a key element in a classification system aimed at distinguishing between three classes composed of HFO events (R, FR) and a mixture of spikes with Fast events (SR) for the dataset (1), and then three classes composed of R, FR, and Rand FR for the data set (2).

The efficiency of the proposed approach in terms of precision, sensitivity, specificity, accuracy, and F-measure has been assessed through numerical experiments using simulated and real iEEG recordings. Our proposed solution addresses this problem by introducing three global models: multiclass SVM, model CNN, and CNN-SVM. The results obtained from these models demonstrate a particularly high level of accuracy.

The results in all models are both encouraging and intriguing. In particular, the GoogleNet-SVM model achieves the highest accuracy rate of 99.63% from the simulated data, while for the real base, it gives an accuracy rate of 94.07%. then our result depended on the data set. This opens up avenues for future research into the detection and localization of epileptic seizures. Future work will focus on the other hybrid algorithms.

ACKNOWLEDGMENT

The authors are grateful to the investigators who shared their iEEG data